\documentstyle[%preprint,
aps%,prb
]{revtex}

\begin{document}

\draft

\title{
Equilibrium and kinetics at the coil--to--globule transition of
star and comb heteropolymers in infinitely dilute solutions
}

\author{
Edward G.~Timoshenko%
\setcounter{footnote}{0}\thanks{Author to 
whom correspondence should be addressed. Phone: +353-1-7162821,
Fax: +353-1-7162127.
Web page: http://darkstar.ucd.ie; 
E-mail: Edward.Timoshenko@ucd.ie}
}
\address{
Theory and Computation Group,
Department of Chemistry, University College Dublin,
Belfield, Dublin~4, Ireland}

\author{Yuri A. Kuznetsov%
\thanks{ E-mail: Yuri.Kuznetsov@ucd.ie }}
\address{Centre for High Performance Computing Applications, 
University College Dublin, Belfield, Dublin~4, Ireland}

\date{\today}

\maketitle

\begin{abstract}
%$\left.\right.$
%\par\noindent
By means of continuous space Monte Carlo simulation we study conformational 
structures formed by star and comb heteropolymers during kinetics of folding 
from the coil to the globule, as well as the corresponding equilibrium states on 
going from the good to the poor solution. Particular examples of combs with 
hydrophobic backbone and hydrophilic side--groups (and vice versa), as well as 
stars with flexible and semi--stiff arms are studied. 
It is interesting to note that 
star--like conformations naturally appear for a comb polymer with a strongly 
hydrophobic backbone. 
We emphasise the crucial difference in the spatial 
distribution of hydrophobic and hydrophilic monomers within the globular 
states for the above mentioned two types of combs. 
In case of stars, the 
non--equilibrium states during kinetics of the coil--to--globule transition 
correspond to formation of localised pearls within flexible arms, whereas
semi--stiff arms prefer to join with each other remaining essentially extended. 
These studies present natural extension of our previous works on the equilibrium and 
kinetic properties of linear and ring heteropolymers based on the Gaussian
self--consistent method and lattice Monte Carlo technique. However, in studying 
polymers with more nontrivial topology continuous space simulation has some 
essential advantages. 
$\left.\right.$
\end{abstract}

\section{Introduction}\label{sec:intro}

From the polymer design perspective three properties of
a polymer are of major importance for the resulting conformational
properties and consequent industrial applications.
These are the degree of stiffness of the chain, the topology of
the polymer, and its composition in terms of monomers involved.
In trying to model, for instance, the structure of 
a natural protein one has to
construct the main chain of a given stiffness and to attach various
side chains, each of which would have required local stiffness,
degree of hydrophobicity and other more specific properties
associated with the nature of the corresponding amino acid residue.
The challenging nature of the kinetics of protein folding would
manifest itself in the shear complexity and diversity of the kinetics
of folding of our hypothetical model polymer, which should  be
rather sensitive on the details of these three types of interactions
involved. 
In the current paper we shall discuss the results from Monte Carlo
simulations in continuous space for the kinetics of
folding of several sufficiently simple and yet already complex types 
of polymers based on a rather generic coarse--grained model
introduced here. These preliminary studies of the model produce a number
of novel transient, and in some cases metastable, kinetic intermediates
and rather complex kinetic pathways of folding. 

Conformational transitions of homopolymers with varying 
degree of flexibility have
been of considerable interest for analytical studies
\cite{Ganazzoli-95,Vasil-97}
and computer simulations \cite{FrenkelStiff,Binder-98} in recent years.
It is well known that folding of a sufficiently stiff
chain results in formation of a torus \cite{TorusFirst} 
under certain conditions. Moreover, the transitions
coil--to--torus and torus--to--globule are discontinuous,
resulting in a rather complex kinetic picture of expansion or collapse,
essentially dependent on the quench depth \cite{Torus}.
The phase diagram of a stiff homopolymer obtained
from the Gaussian variational method \cite{Torus}
is found to contain phases
of the coil, a spherical (or slightly aspherical) globule in the region of
either a low stiffness or a strong monomer attraction, and a number of
toroidal phases characterised by distinct winding number in between.
The collapse transition changes its behaviour from continuous to
discontinuous starting from some value of the stiffness.
This distinction is quite dramatic for the kinetics of collapse.
While the kinetics of the coil--to--globule transition for a flexible
homopolymer proceeds via formation of a necklace of
locally collapsed sub--globules or `pearls' \cite{GscKinet} rather
than via a sausage mechanism\cite{deGennesKinet}, the pearls are not seen
in the semi--flexible case where hairpins with abrupt U-turns
and indeed sausages, often also corresponding to metastable states,
would be typical transient kinetic intermediates \cite{Torus}.

There is also a great deal of interest in  the properties of
amphiphilic heteropolymers. Even in the flexible case these exhibit
a variety of complex conformational transitions
upon a temperature or pH variation \cite{Dill,Halperin,Israel}.
Apart from the stiffness and the sequence of monomers, one can
also play with various polymer topologies \cite{Rouault}.
Indeed, random branching of polymers is quite common during synthesis.
By a carefully controlled polymerisation procedure  one can obtain 
regular stars, which presented a nice object for theoretical study due to
their symmetry and simplicity lately \cite{Ganazzoli-95,CopStar}.
Star polymers are believed to find a number of future applications for
coating, as additives and possibly in drug delivery systems related
to their low viscosity and other interesting properties.

Speaking of computer simulations,
star polymers, for instance,  (see e.g. Refs. \cite{Grest,Freire})
have been studied by a number of techniques,
ranging from Monte Carlo, both on and off lattice,
to Brownian and Molecular Dynamics \cite{Freire}.
While dynamic Monte Carlo is one of perhaps most efficient
approaches in terms of accessible time--scales, on a a lattice
it has a number of obvious limitations.
Indeed, the core
monomer on a lattice has a great difficulty to move at all due to its high
coordination number (equal to the number of arms). The
resulting rejection of all attempted moves
for some kind of motions is called a quasi--nonergodicity, a problem
which is hard to deal with even for linear copolymers. If such a
situation occurs, it practically means that the results of the
simulation may no longer be trusted.
A number of complicated non--local moves involving the core
monomer have been suggested to alleviate this problem \cite{Molina}.
However, such moves are not permitted in case of Dynamic Monte Carlo needed
for study of, for instance, metastable states, moreover they may present
problems due to subtle topological effects for truly non--phantom
chains \cite{TopologyMC}.

Referring to our recent paper \cite{CopStar} for
examples of applications to diblock copolymer star,
we devote the present work to investigating the kinetics
of the coil--to--globule transition for flexible and semi--flexible
homopolymer stars and diblock copolymer combs with the hydrophobic
backbone (or H-backbone comb briefly) with hydrophilic (polar) side--arms and
vice versa (or P-backbone comb).

\section{Results}
\label{sec:results}

The mathematical model for a generic coarse--grained polymer is based
upon the following Hamiltonian (energy functional):
\begin{eqnarray}
H & = & \frac{k_B T}{2\ell^2}  \sum_{i\sim j} \kappa_{ij} ({\bf X}_i - {\bf X}_{j})^2
      + \frac{k_B T }{2\ell^2} \sum_{i\approx j \approx k} \lambda_{ijk}  ({\bf X}_{i} +
                              {\bf X}_{k} - 2{\bf X}_j)^2  \nonumber \\
  & & + \frac{1}{2} \sum_{ij,\ i\not= j} V_{ij} (|{\bf X}_i - {\bf X}_j|).
\label{cmc:hamil}
\end{eqnarray}
Here the first term presents the connectivity structure of the polymer with
harmonic springs of a given strength $\kappa_{ij}$ introduced between
any two connected monomers (which is denoted as $i\sim j$).
The second term presents the bending energy penalty given by the
square of the local curvature with a characteristic stiffness $\lambda_{ijk}$
between any three consecutively connected monomers (which is
denoted as $i\approx j \approx k$) in the form of the Kratky--Harris--Hearst
term \cite{Kratky,Harris}.
Finally, the third term presents pair--wise non--bonded interactions
between monomers such as van der Waals forces and so on.
In a simple model we can adopt the Lennard--Jones form for the
shape of the potential,
\begin{equation}
V_{ij}(r) = \left\{
\begin{array}{ll}
+\infty, &  r < d \\
V_{ij}^0 \left( \left( \frac{d}{r}\right)^{12}
- \left( \frac{d}{r} \right)^{6} \right), &  r > d
\end{array}
\right.,
\end{equation}
where there is also a hard core part with the monomer diameter $d$
(below this is taken to be equal to the statistical length $\ell$,
i.e $d=\ell$). Finally, we shall restrict ourselves to a binary
heteropolymer with hydrophilic and hydrophobic units, so that the
potential coefficients can be further parametrised as \cite{ConfTra},
\begin{equation}
V_{ij}^0=\frac{V_i+V_j}{2}, \quad V_i=V_0(1-\sigma_i), \quad \sigma_i = \pm 1.
\end{equation}
For details of the adopted Monte Carlo technique we refer to our
paper Ref. \cite{Torus}.

Our first subject is the collapse kinetics of a flexible ($\lambda=0$) 
homopolymer star after a quench from the purely repulsive good
solvent condition ($V_0=0$) to the poor solvent ($V_0=5$). Here and
below $k_B T$ units are used for the energy and $\ell$ units for
the distance.
From Fig.~\ref{fig:flstar}a one can see that the process involves
formation of locally collapsed clusters along the chain similarly
to the case of an open or ring chain. However, due to the connectivity
structure of the star, the largest cluster is formed around the core
monomer and the arms ends seem to belong to fairly large clusters.
Also, it is interesting to note that some of the clusters involve
monomers from more than one arm (see e.g. on the bottom right of the figure).
Further, the ends of arms clusters are being pulled towards the growing
core cluster, so that the total number of clusters and free arms
rapidly decrease (Fig.~\ref{fig:flstar}b). Finally, an almost spherical
globule will form and continue to optimise its shape and internal
structure until reaching the final equilibrium exponentially slowly 
\cite{GscKinet}.

In Figs.~\ref{fig:rdstar}, similarly, we examine the kinetic process for a
semi--flexible homopolymer star with $\lambda=5$. Clearly, as in the case
of a semi--flexible rings in Ref. \cite{Torus} formation of the localised
clusters is no longer possible due to a large energetic penalty.
Instead, as in Fig.~\ref{fig:rdstar}a some of the arms come together
forming a kind of fibrils, some of which still have free outstretched
arms. This tree--like structure further experiences the process of
fibrils growth with their width increasing and the number and
lengths decreasing (Figs.~\ref{fig:rdstar}b and \ref{fig:rdstar}c).
The overall compaction process is much slower compared to the flexible
case and finally a rather long--lived sausage--like object would remain.
It is interesting to point out that quite often these kinetically arrested
states would have one or a few pieces of single arms sticking out of
the the main irregularly shaped body as in Figs.~\ref{fig:rdstar}c
and \ref{fig:rdstar}d. These curious looking conformations are rather
typical for a semi--flexible chain and should in principle 
be seen by Electron Microscopy in experiments involving long 
semi--stiff star polymers. This effect is clearly related to the topology
of a star since similar conformations are not present for a ring polymer
\cite{Torus}.

Now let us turn our attention to flexible comb copolymers. We analyse
the simple case in which the backbone is constructed of one type of
monomers, while the side--arms of another, one of which is
hydrophobic (H) and another hydrophilic (P for polar), so that the
total numbers of monomers of both kinds are equal. In particular
we shall analyse the situation in which in the final state
$V^0_{PP}=0$ (pure repulsion), $V^0_{HH}=5$ and $V^0_{HP}=2.5$,
corresponding to a near ideal regime for the H-P monomer interaction.
In Figs.~\ref{fig:eqlcomb} we present the characteristic equilibrium
conformations for the cases of the H-backbone (Fig.~\ref{fig:eqlcomb}a)
and the P-backbone (Fig.~\ref{fig:eqlcomb}b). The former case pretty
much reminds the situation for a OUTER-P star \cite{CopStar}, so
that in Fig.~\ref{fig:eqlcomb}a we see the collapsed hydrophobic backbone
in the centre with three outstretched hydrophilic side--groups
(the forth one is pointing in the perpendicular to the plane direction
and thus is not visible). This object can be viewed as a micellar
star. The reversed case of the P-backbone again produces a micellar
object with a single collapsed core of hydrophobic side--groups
and extended hydrophilic backbone chain looping around the core.
This is a much more compact object compared to the H-backbone case
and it is in many ways similar to the collapsed OUTER-H star in
Ref. \cite{CopStar}. Also, the H-core in Fig.~\ref{fig:eqlcomb}b
is clearly much less exposed to unfavourable contacts with the
solvent molecules than in Fig.~\ref{fig:eqlcomb}a.

Next, let us also discuss briefly the transient kinetic states
after a quench from the good solvent condition (all $V^0_{ij}=0$) to
the above discussed final equilibrium values in both cases.
In Fig.~\ref{fig:kincomb}a one can see the formation of locally
collapsed clusters along the hydrophobic back backbone, which further
grow and merge with each other until producing a single globular
core with the hydrophilic side--arms remaining extended and outstretched
throughout. This mechanism is akin to the case of an open homopolymer
collapse except that perhaps the mobility of the growing clusters is
somewhat reduced by the presence of the passive side--arms.
Finally, in Fig.~\ref{fig:kincomb}b we present a typical non--equilibrium
conformation for the P-backbone comb. Here, the folding process starts
very quickly by a full collapse of all individual hydrophobic side--arms,
which remain farther apart from each other along the hydrophilic backbone.
This arrangement results in a rather long--lived  metastable state,
which has a rather distinct structure from the final equilibrium
state depicted in Fig.~\ref{fig:eqlcomb}b. An example of a similar
metastable state in kinetics was also seen for an OUTER-H star in
Ref. \cite{CopStar}, where we have also seen that a much deeper
quench would render such a state unstable turning the nucleation
regime into that of spinodal decomposition.

%%%%%%%%%%%%%%%%%%%%%%%%%%%%%%%%%%%%%%%%%%%%%%%%%%%%%%%%%%%%%%%%%%%%%%%%

\section{Conclusions}

In this work we have presented some preliminary results on the
studies of the equilibrium and kinetics of conformational changes in
flexible and semi--flexible homo- and block copolymer stars and combs.
We have seen a number of unusual conformational states and distinct
folding pathways strongly dependent on the macromolecular architecture.

Study of the homopolymer collapse kinetics in stars of varying 
degree of flexibility has elucidated that the persistent length of
the chain does play a major role in determining whether the necklace
of clusters or the fibril growth mechanism of folding is realised.
This was also true for simpler topological objects such as an open
chain or a ring \cite{GscKinet,Torus}. However, there are some
topology induced and previously unseen features such as for instance
a `monkey and its tail' conformation in Fig. \ref{fig:rdstar}d.
Further increase of the stiffness may lead to some novel
equilibrium phases  and a complex resulting phase diagram 
in analogy with multiple toroidal states having distinct winding numbers
for a ring homopolymer \cite{Torus}.

Also, it is interesting that a number of similarities with the previously
studied case of diblock star copolymers \cite{CopStar} has been seen here
for comb polymers. It is quite likely that many of these features
would be also retained for more complex regularly branched
dendrimers \cite{Tomalia}, which are currently being scrutinised. 

It would be important to investigate attendant kinetic laws and estimate
characteristic time scales involved as we have done for the homopolymer
kinetics \cite{GscKinet,Torus,ConfTra}. In doing so, it would be
tempting to use several alternative simulation techniques 
such as a coarse--grained Newton and Langevin Dynamics and
an analytical technique based on the Gaussian self--consistent
treatment \cite{ConfTra} for being able to obtain the most
complete picture in this challenging and important problem.

Finally, we believe that modelling chains with distinct types 
of side--groups is necessary for more realistic description of the kinetics of 
protein folding. Undoubtedly, a simple heteropolymer model of proteins
does miss many of the important interactions and details present in 
real polypeptides, and a proper account for side--groups and charges
of the residues is equally important as the hydrophobic force in
driving the folding.

\acknowledgments
The authors acknowledge interesting discussions with Professor
F.~Ganazzoli, Professor H.~Orland, Professor T.~Garel and
Ronan Connolly.
This work was supported by grant FR/2000/019 from Enterprise Ireland
and UCD President's research award.
We also acknowledge the support of the Centre for High Performance
Computing Applications, UCD.

%%%%%%%%%%%%%%%%%%%%%%%%%%%%%%%%%%%%%%%%%%%%%%%%%%%%%%%%%%%%%%%%%%%%

%%%%%%%%%%%%%%%%%%%%%%%%%%%%%%%%%%%%%%%%%%%%%%%%%%%%%%%%%%%%%%%%%%%%%%%%

\newpage

%\section*{Figure captions}

%%%%%%%%%%%%%%%%%%%%%%%%  Fig. 1 (a,b)  %%%%%%%%%%%%%%%%%%%%%%%
\begin{figure}
\caption{
Snapshots of typical non--equilibrium conformations of a flexible star
homopolymer during collapse kinetics.
This and consequent figures are obtained from the continuous space Monte Carlo 
simulation.
Here the star consists of $f = 12$ arms with $N/f = 50$ monomers in each arm.
Figs. a and b are obtained after $t = 20,000$ and $t = 100,000$ of
Monte Carlo sweeps (MCS) respectively (1 MCS is equal to $N+1$ attempted
Monte Carlo moves).
The kinetics is analysed after the quench of the
two--body interaction parameter $V^0 = 0 \rightarrow 5$ (in $k_B T$ units).
}\label{fig:flstar}
\end{figure}

%%%%%%%%%%%%%%%%%%%%%%%%  Fig. 2 (a-d)  %%%%%%%%%%%%%%%%%%%%%%%
\begin{figure}
\caption{
Snapshots of typical non--equilibrium conformations of a semi--flexible
($\lambda = 5\ell$) homopolymer star during collapse kinetics.
Figs a--c are obtained for the same initial condition after
$t = 35,000$, $t = 350,000$ and $t = 10^6$ MCS respectively. 
Fig.~d is obtained from another initial condition after
$t = 10^6$ MCS.
Star composition and quench depth here are the same as 
in Fig.~\ref{fig:flstar}.
}\label{fig:rdstar}
\end{figure}

%%%%%%%%%%%%%%%%%%%%%%%%  Fig. 3 (a,b)  %%%%%%%%%%%%%%%%%%%%%%%
\begin{figure}
\caption{
Snapshots of typical equilibrium conformations of a flexible
comb copolymer.
Figs.~a and b correspond to the H-backbone and P-backbone combs respectively.
Here the number of monomers is $N = 288$, the number of side--chains is $f = 4$,
and the two--body interaction parameters are 
$(V^0_{PP}, V^0_{HP}, V^0_{HH}) = (0, 2.5, 5)$ (in $k_B T$ units).
Black and light grey circles correspond to hydrophobic and hydrophilic
monomers respectively.
}\label{fig:eqlcomb}
\end{figure}

%%%%%%%%%%%%%%%%%%%%%%%%  Fig. 4 (a,b)  %%%%%%%%%%%%%%%%%%%%%%%
\begin{figure}
\caption{
Snapshots of typical non--equilibrium conformations of a flexible
comb copolymer during folding kinetics.
Figs.~a and b correspond to the H-backbone and P-backbone combs respectively.
Here the quench depth is
$(V^0_{PP}, V^0_{HP}, V^0_{HH}) = (0,0,0) \rightarrow (0, 2.5, 5)$
(in $k_B T$ units),
while the rest of parameters are as in Fig.~\ref{fig:eqlcomb}.
}\label{fig:kincomb}
\end{figure}

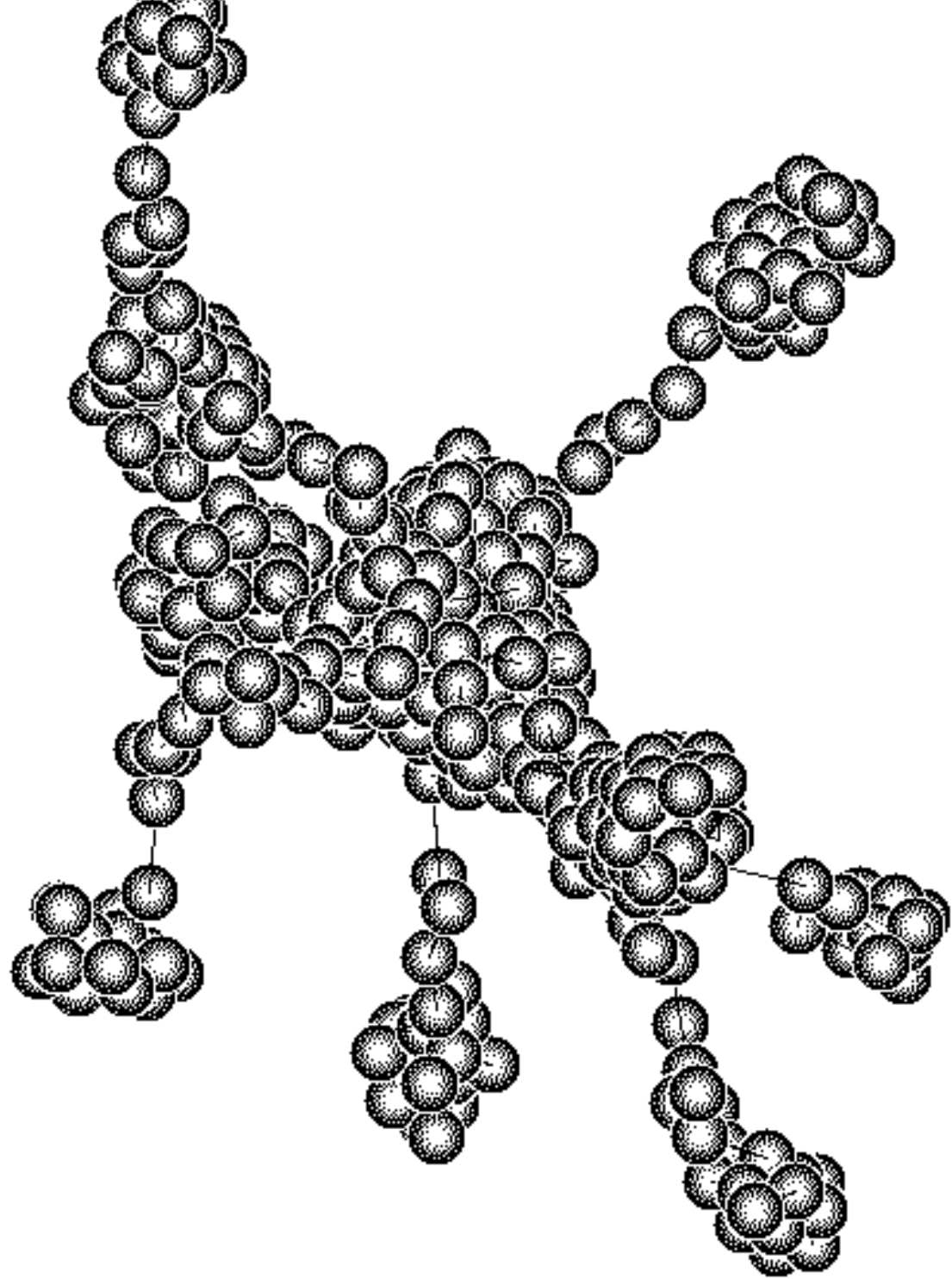
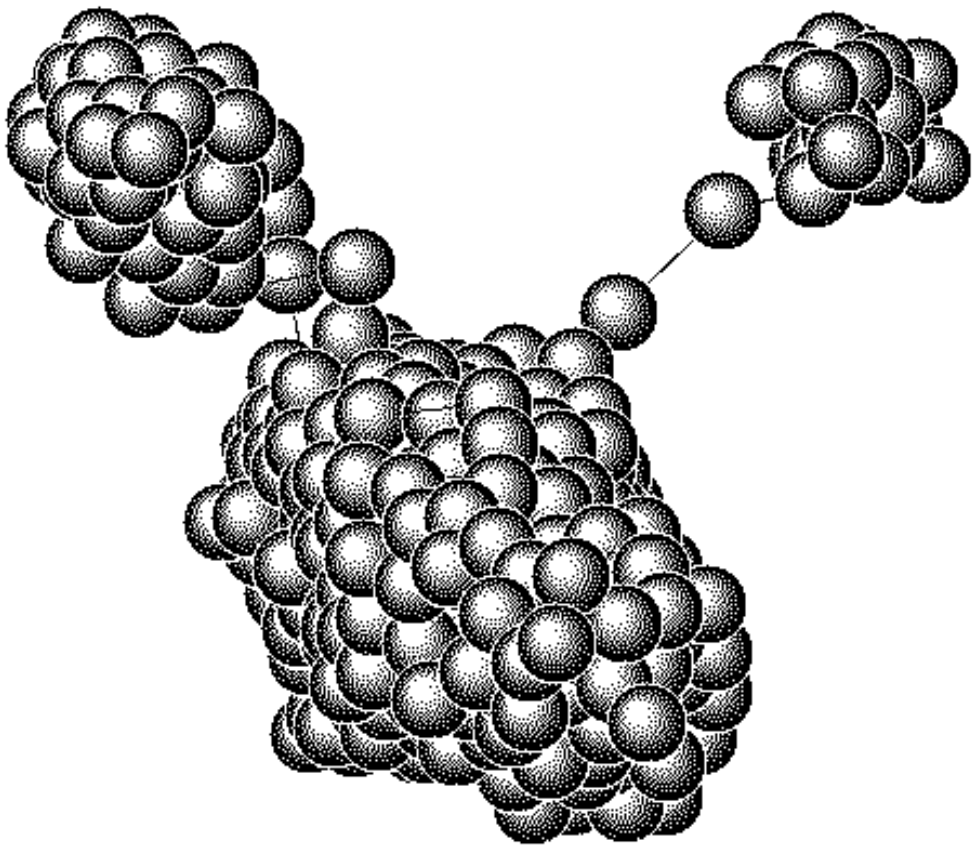
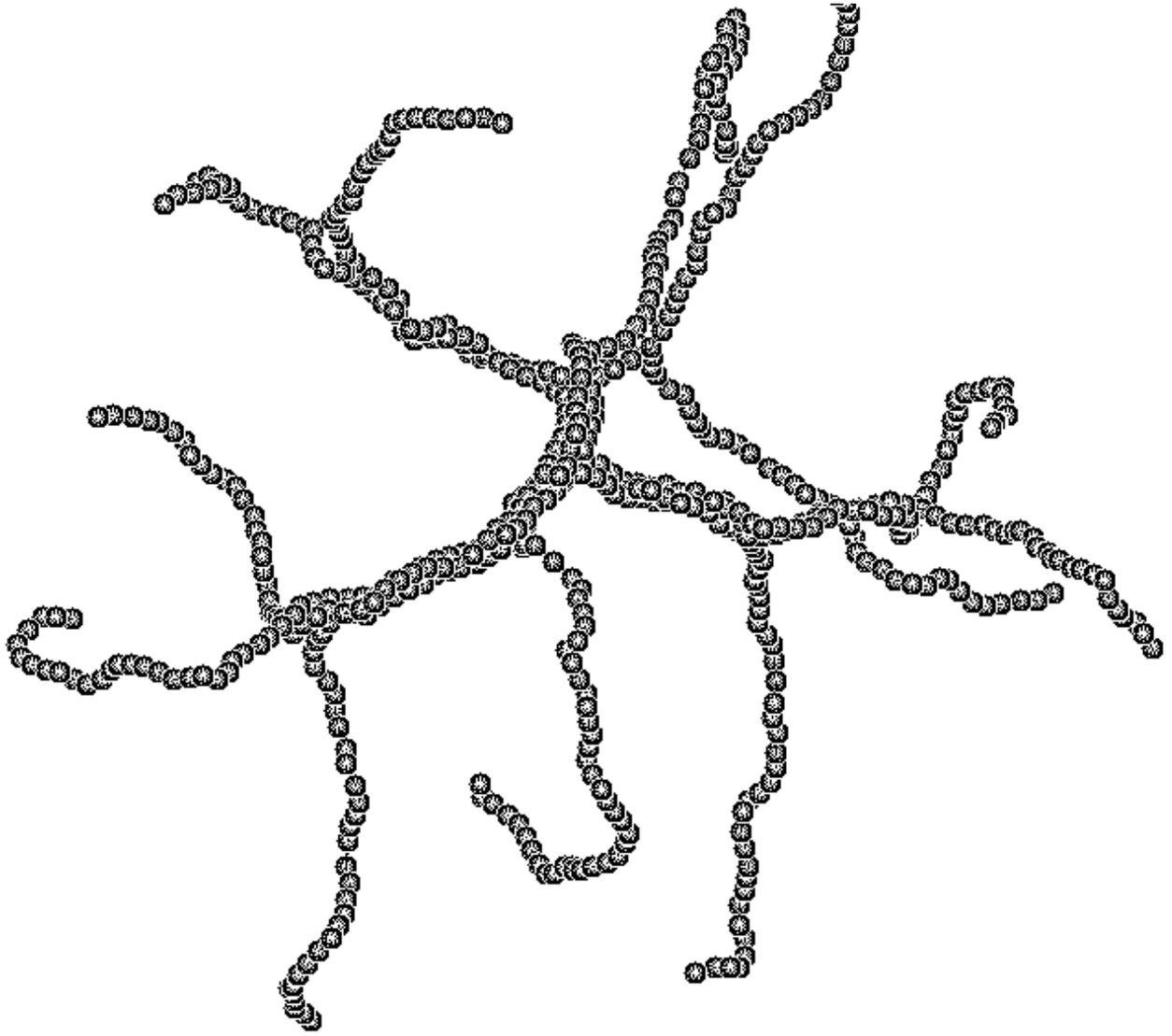
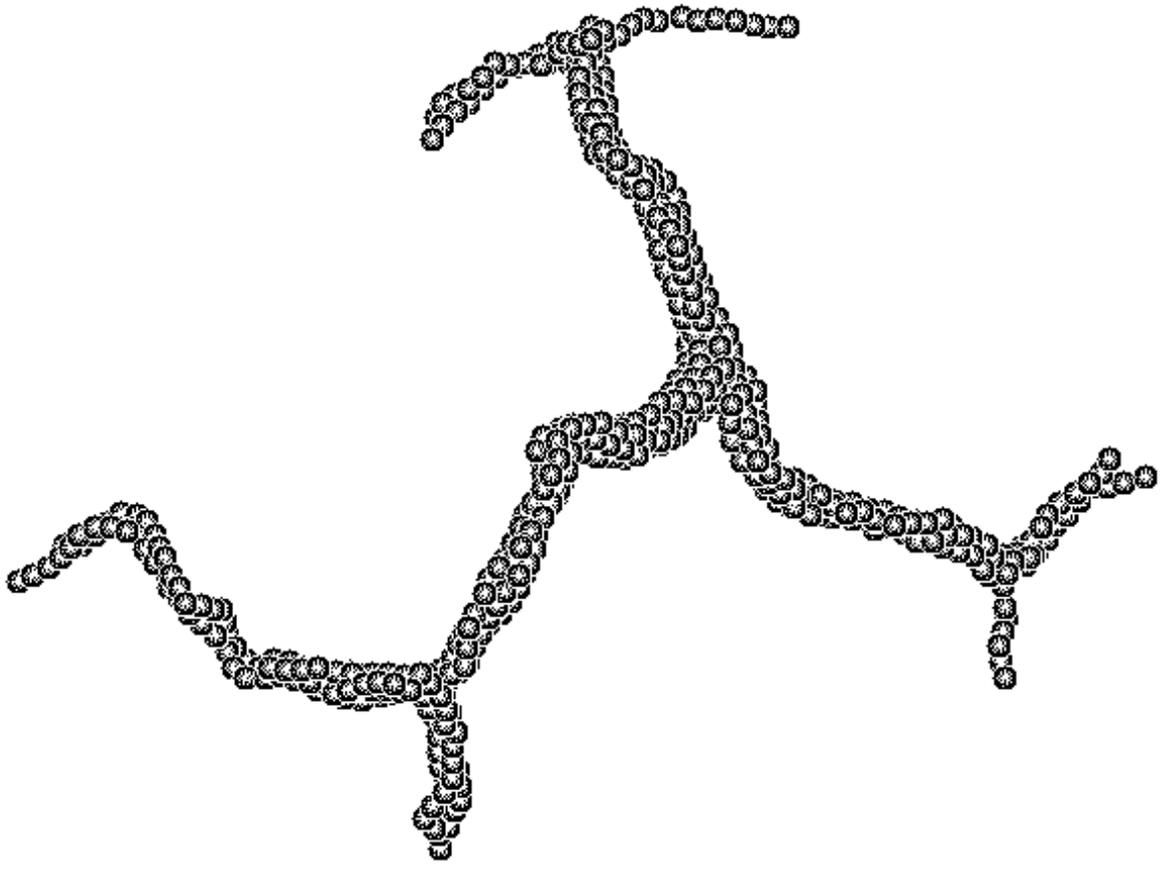
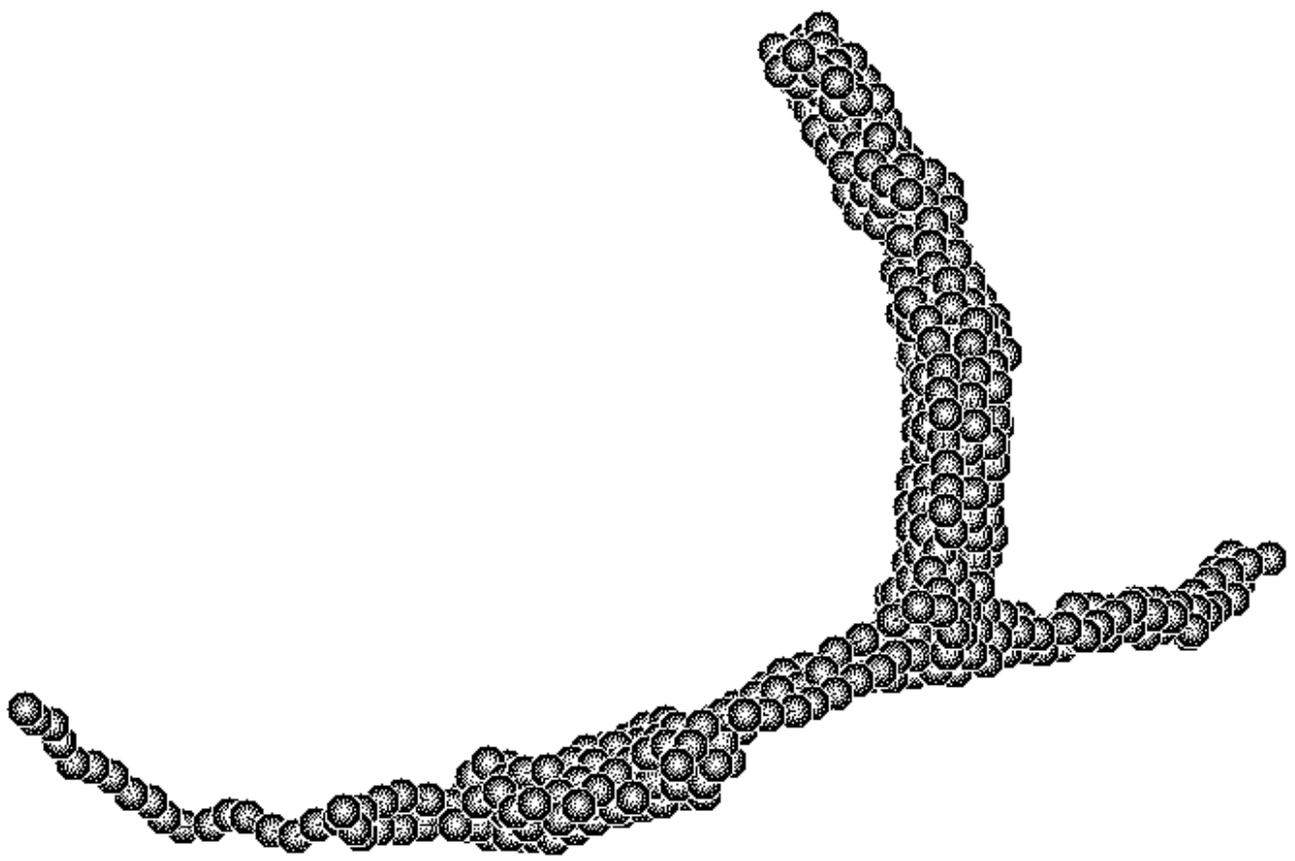
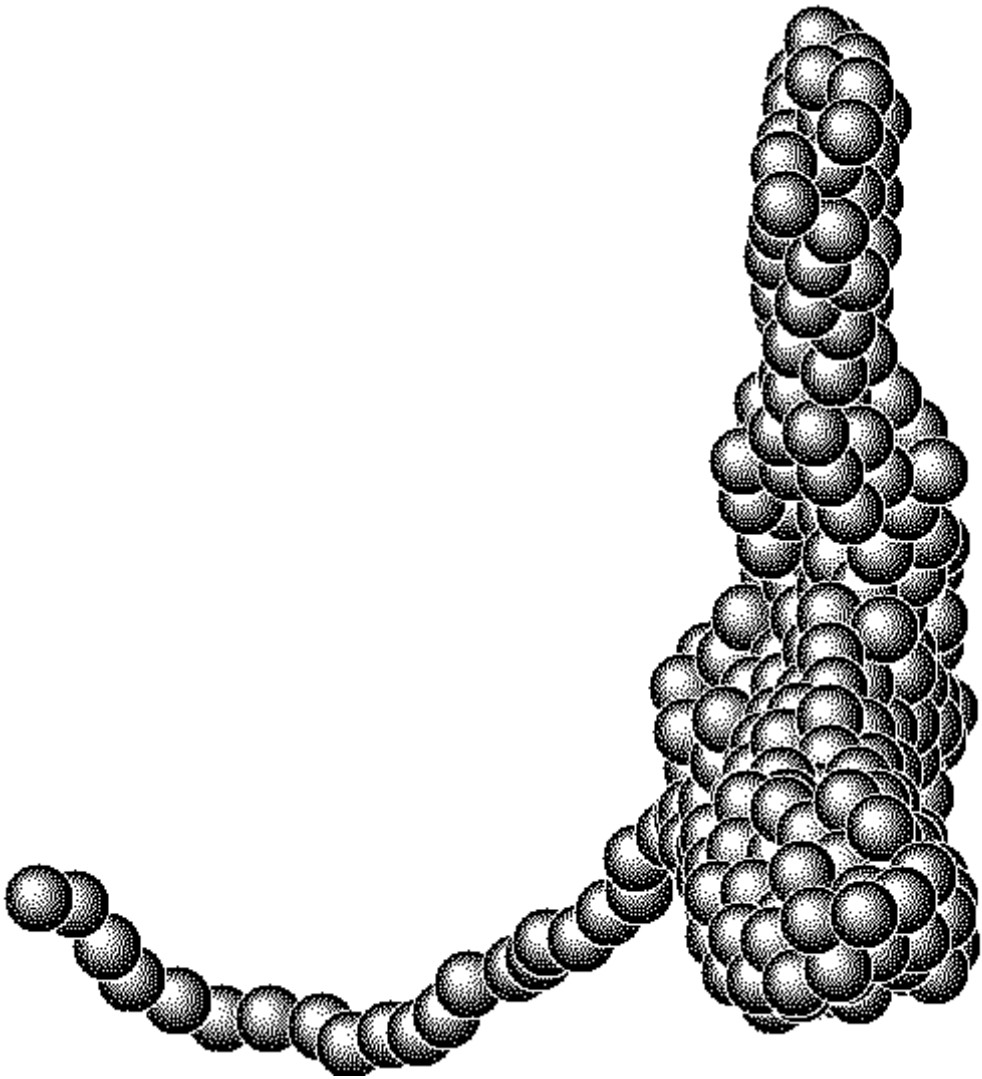
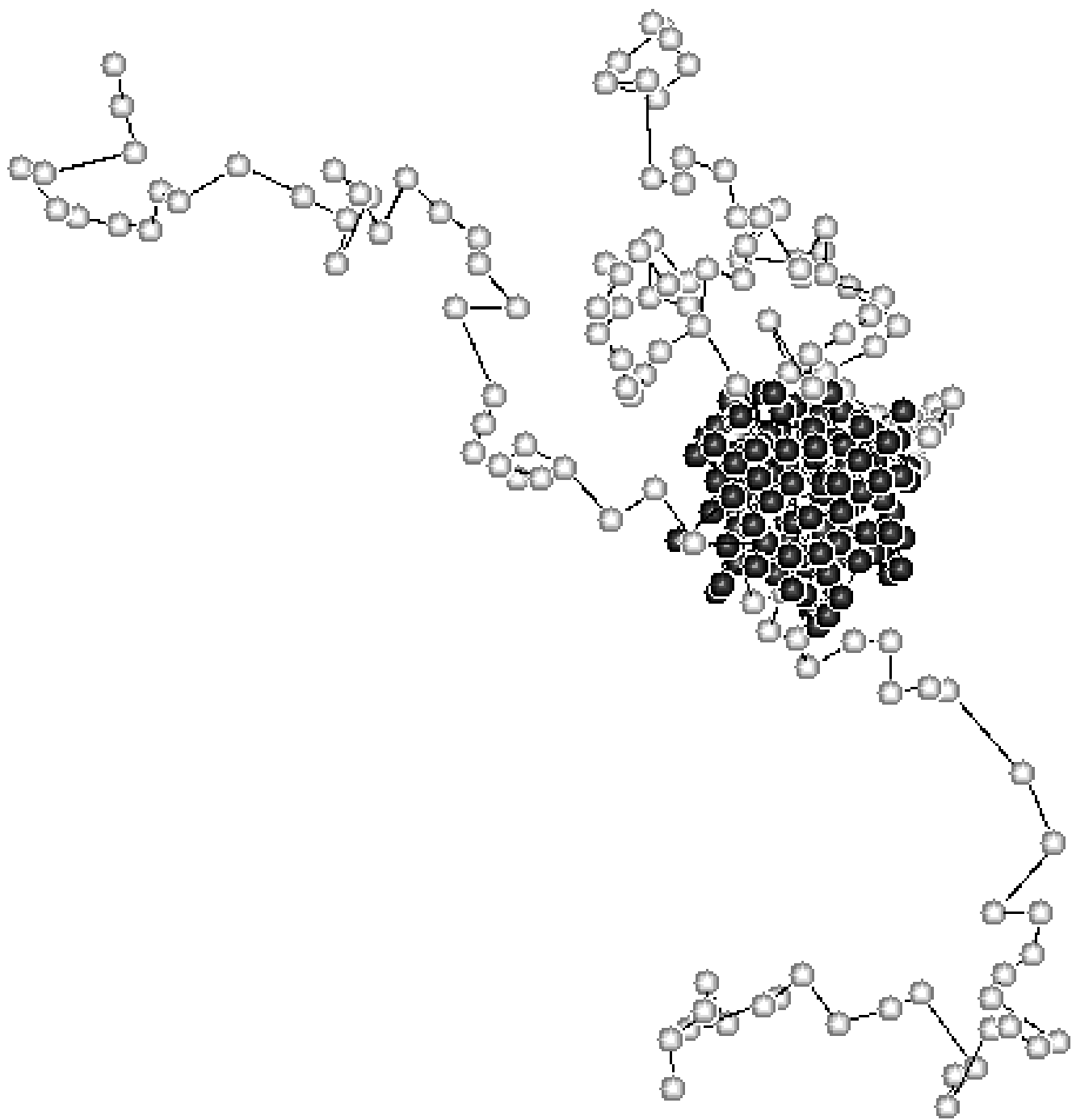
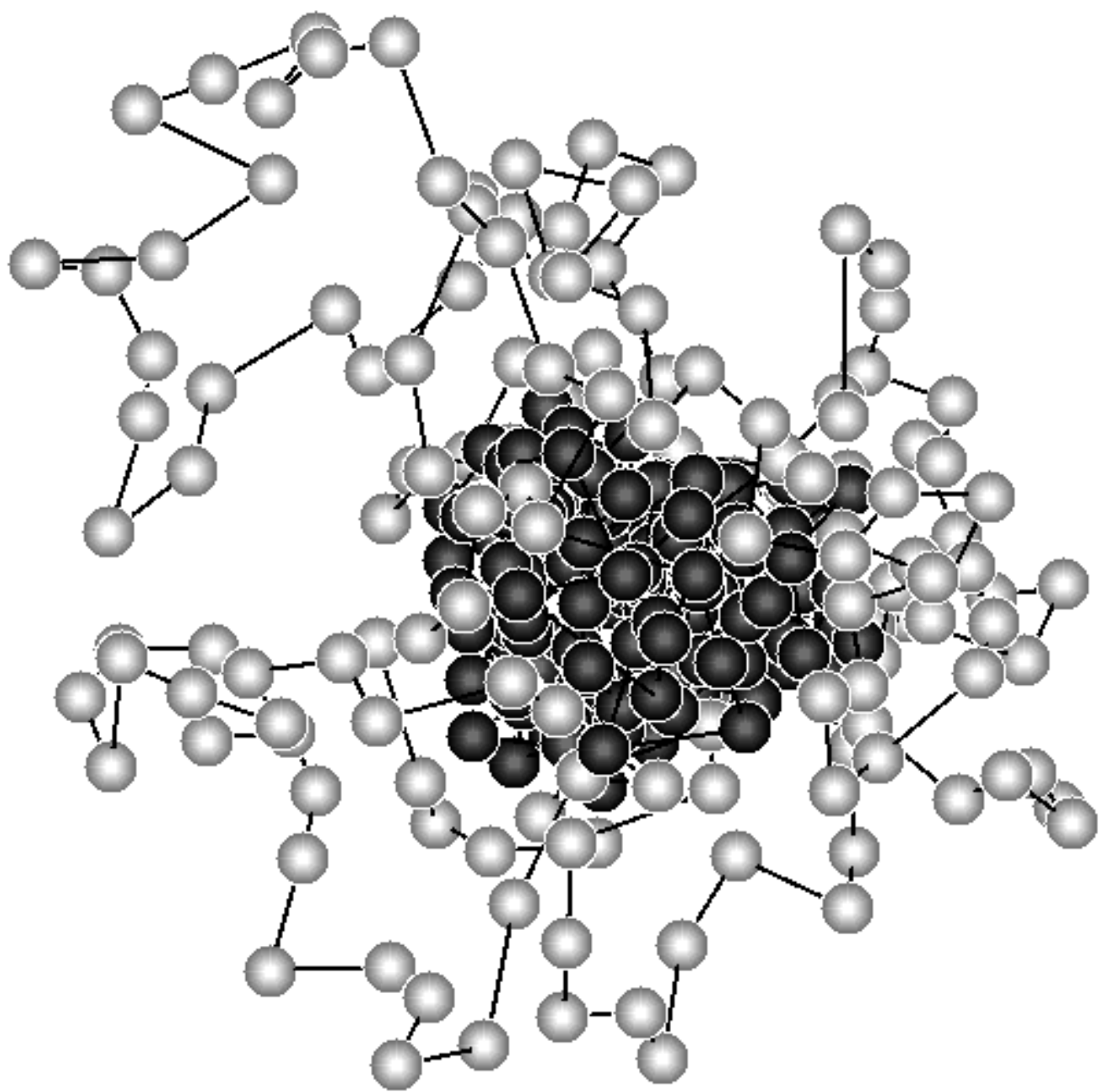
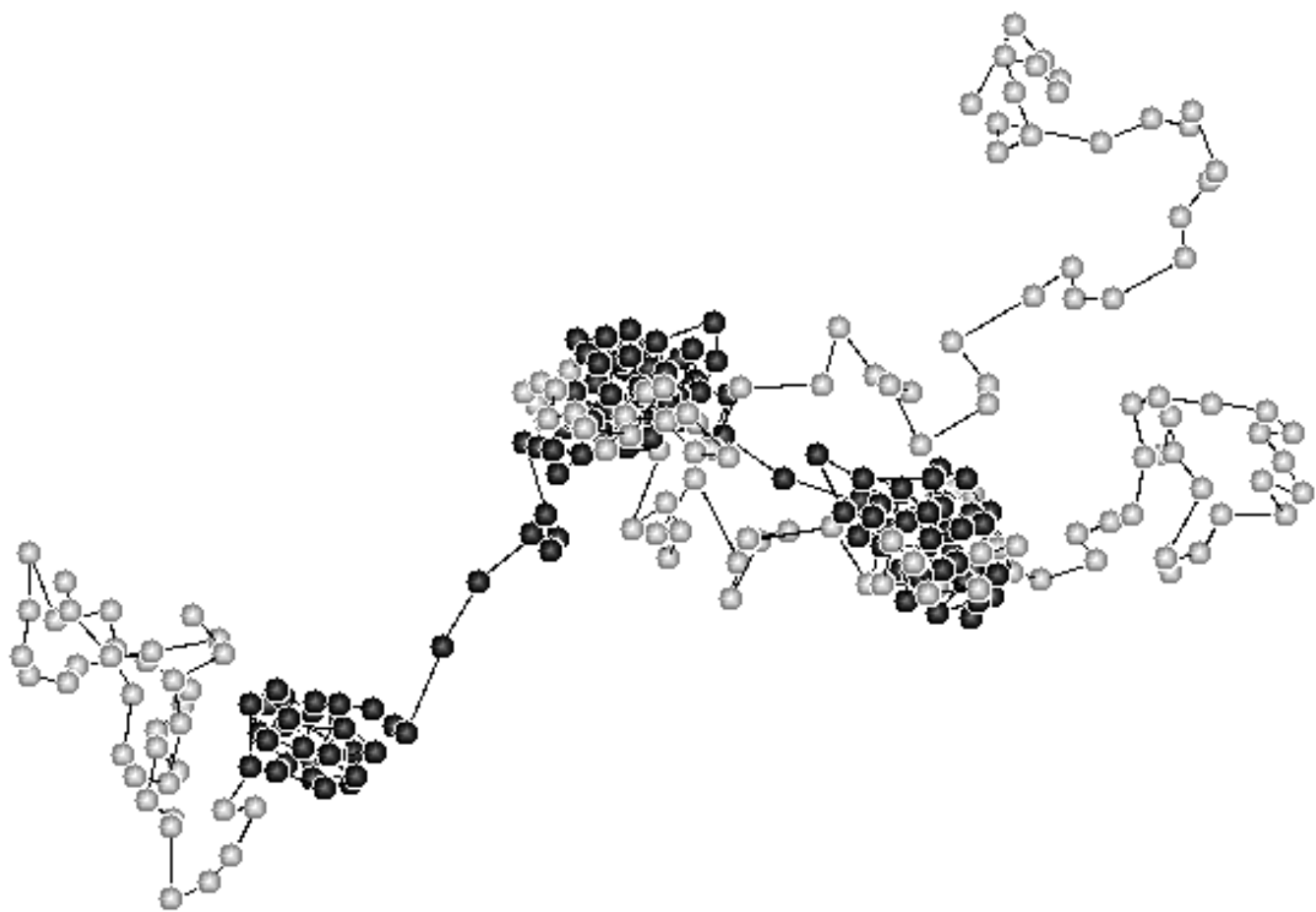
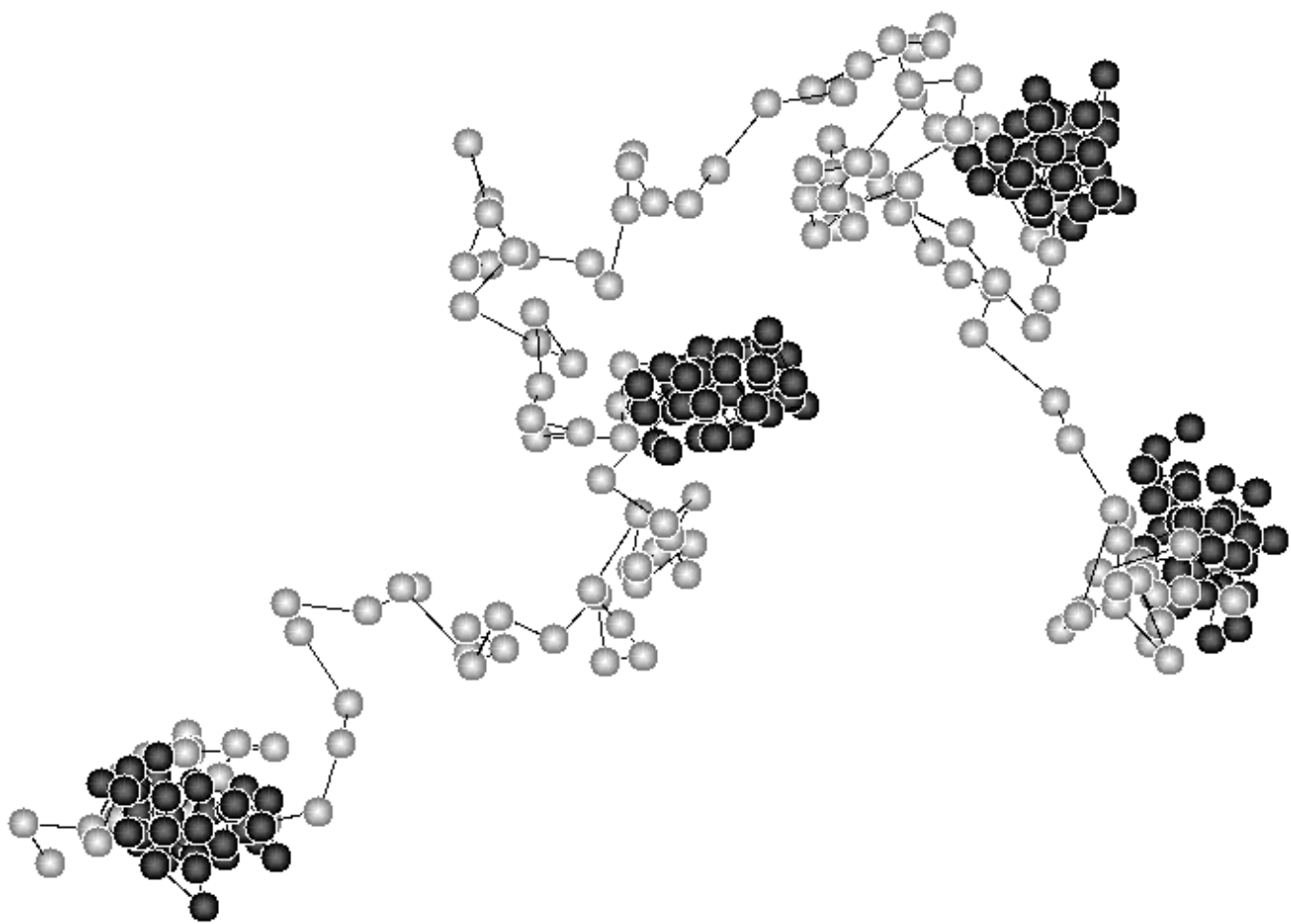
\end{document}